# Incoherent illumination for motion-based imaging through thick scattering medium

Zhao Wang[1,+], Rui Ma[2,+], Mingzhu She[1,+], Anda Shi[1], Weili Zhang[1,*]

[1] Fiber Optics Research Centre, School of Information and Communication Engineering, University of Electronic Science & Technology of China, Chengdu, 611731, China

[2] Institute of Advanced Study, Shenzhen University, Shenzhen 518060, China

[+] These authors contributed equally to this work

*Corresponding author: wl_zhang@uestc.edu.cn*

**Abstract:** Object-motion-based speckle correlation can recover hidden objects from any inhomogeneous medium, which takes advantage of the inherent connection that the cross-correlation between speckle patterns can reflect the object's autocorrelation, providing a route for imaging through or inside thick scattering media. However, once the object is phase-modulated, the above-mentioned relation will not be satisfied under coherent illumination, and the objects can not be recovered using the existing approaches. Here, we propose an incoherent illumination method for object-motion-based imaging. Theoretical analysis and experimental results show that the cross-correlation between the object-motion-based speckle patterns can be directly used to represent the intensity autocorrelation of the object, making it possible to recover hidden objects regardless of whether the object is phase-modulated or not. Moreover, the proposed approach has a lower root-mean-square error for extracting the autocorrelation patterns of the hidden object. The proposed imaging mechanism blazes a way of imaging moving objects with scattering-induced or intrinsic phase profile, which is in favor of complex imaging scenarios such as inhomogeneous object imaging, deep tissue imaging, and passive lighting scattering imaging.

Introduction:

The phenomenon of scattering widely exists in everyday life, which generally reduces the signal-to-noise ratio of imaging systems or even submerges useful object information in speckle noise. Traditional approaches separate ballistic photons (i.e., useful information) from scattering photons using methods like time gating or spatial gating [1-5]. Unfortunately, the intensity of the ballistic photons decreases exponentially with depth, while the scattering photons dominate the optical field for deeper imaging depths [6]. New technologies have emerged to extract effective information from strongly diffused photons, such as wavefront shaping [7-9], transmission matrix measurement [10-12], and memory effect-based speckle-correlated imaging [13-16]. However, these methods tend to become less effective as the thickness of the scattering medium increases, due to the rising number of spatial independent modes. Object-motion-based speckle correlation (OMSC) can be used to recover hidden objects regardless of the shape, material, or thickness of the scattering medium, which is promising to image auto-moving objects in turbid environments, such as stars or blood cells [17-21].

Recently, the theory and technology of OMSC under coherent illumination (OMSC-CI) have been enriched. In 2014, K. J. Webb et al. introduced the concept of OMSC-CI and demonstrated that this method can extract the autocorrelation of binary objects regardless of the thickness of the scattering medium [17]. In 2016, they extended the OMSC-CI approach to the scenario where objects are embedded within a scattering medium [19]. In 2022, T. Shi et al. imaged moving objects that were obscured by a random corridor through the subspace reduced OMSC-CI [21]. These advances are conducive to addressing the restriction of object reconstruction in imaging where a complex scattering medium is involved.

For OMSC imaging through or inside complex scattering media, current research mainly focuses on the hidden object with a binary pattern and neglects spatial phase modulation. However, the phase-modulated object also exists in quite a few practical imaging systems, which generally can be divided into two categories. One is extrinsic phase modulation generated from coherent illuminating light passing through the inhomogeneous medium where the object is embedded, and the other one is intrinsic phase modulation originating from the inhomogeneous surface or index distribution of the object. Once the object is phase modulated, the OMSC-CI can no longer reflect the object's autocorrelation since the intensity cross-correlation between speckle patterns is the square of an irreversible function of the object's field under coherent illumination. Therefore, the existing method of OMSC-CI would fail to reconstruct the phase-modulated object.

Here, OMSC under incoherent illumination (OMSC-II) is proposed and particularly analyzed. Firstly, the influence of spatial coherence on OMSC is thoroughly investigated based on wave optics. Both theoretical analysis and experimental results show that the OMSC-II can directly represent the intensity autocorrelation of the hidden object, while the OMSC-CI represents the squared module of field autocorrelation of the object. On this basis, our method is insensitive to the phase distribution of the object and can extract the intensity autocorrelation of a phase-modulated object from the cross-correlation of speckle patterns, while the effective information is almost lost when using the former method. Moreover, the OMSC-II approach has a lower error in extracting autocorrelation of the object, even in binary modulated object imaging. Thus, the proposed method shows strong applicability and promising potential for imaging of phase-modulated objects, deep tissue fluorescent labeling imaging, and passive lighting scattering imaging.

**1. Influence of spatial coherence on OMSC.**

The model of imaging through a thick scattering medium under incoherent illumination is depicted in Fig. 1(a). Spatially incoherent illumination is generated by passing a laser through a fast-rotating ground glass. The light is then used to illuminate a moving object, i.e., a real-time pattern generated by a digital mirror device (DMD). In our work, it is worth emphasizing that as a directly illuminated object, the DMD here only provides intensity modulation without phase and gray-scale modulation. Thus, intensity and field autocorrelations of the binary object are equivalent. The images of the moving object at different coordinate positions and autocorrelation patterns of the object are given in Fig. 1(b). Similarly, Fig. 1(c) corresponds to the case of coherent illumination, wherein the laser is used directly to illuminate the object without spatial decoherence operation (i.e., passing through a rotating ground glass).

Speckle patterns are formed after an object passes through a thick scattering medium. For spatially incoherent illumination, each speckle pattern, $I$, received on the camera is the intensity superposition of point spread functions (PSFs) formed by different points of an object passing through the scattering medium, which can be expressed as (details as shown in Supplementary Note 1)

$$I(r_c; r_o) = \int I_O(r_o) I_{psf}(r_c; r_o) dr_o \qquad (1)$$

Parameter $r$ denotes the transverse spatial coordinate in the $x$ and $y$ plane with subscripts '$o$' and '$c$' representing the object and camera plane respectively. $I_O$ is the intensity pattern of the object which is equal to the squared modulus of the field, $E_O$, (i.e., $I_O(r_o) = |E_O(r_o)|^2$). $I_{psf}$ is the PSF of the scattering imaging system (i.e., speckle patterns captured by the camera when only one point of the object plane is lighted). It is worth emphasizing that unlike previous speckle-correlated imaging (SCI) based on optical memory effects (OME) [22], the PSF no longer has translation invariance due to the strong and heavy scattering of thick scattering media. Thus, the OME-based SCI will fail to calculate the autocorrelation of hidden objects through a single speckle pattern, while the OMSC can be used to recover hidden objects.

When the object is moved to different spatial coordinates, it will generate speckle patterns accordingly on the camera, e.g., the speckle patterns shown in Figs. 1(d) (incoherent illumination) and 1(f) (coherent illumination) respectively correspond to different displacements of the object shown in Fig. 1(b). Cross-correlation ($C$) between the speckle patterns is termed OMSC, which is calculated as follows

$$C(\Delta r) = \frac{\langle (I(r_c;r_o) - \langle I(r_c;r_o) \rangle) \cdot ((I(r_c;r_o + \Delta r) - \langle I(r_c;r_o + \Delta r) \rangle) \rangle}{\sigma_{I(r_c;r_o)} \cdot \sigma_{I(r_c;r_o + \Delta r)}} \quad (2)$$

where $\Delta r$ represents the displacement of the object, $\sigma$ is the standard deviation of the speckle, and $\langle \cdots \rangle$ denotes spatial averaging operation. Substituting Eq. (1) into Eq. (2), we have (details as shown in Supplementary Note 1)

$$C_{II}(\Delta r) = \frac{\int I_O(r_o) I_O(r_o + \Delta r) dr_o}{\int I_O(r_o)^2 dr_o} \quad (3)$$

The subscript '$II$' represents incoherent illumination. The right side of Eq. (3) is the intensity autocorrelation of the object. It is worth emphasizing that Eq. (3) is different from the result of previous coherent illumination methods where the cross-correlation between speckle patterns (Eq. (2)) is equal to the squared modulus of the autocorrelation of the object's field [17, 21], which reads

$$C_{CI}(\Delta r) = \left| \frac{\int E_O(r_o) E_O^*(r_o + \Delta r) dr_o}{\int |E_O(r_o)|^2 dr_o} \right|^2 \quad (4)$$

The subscript '$CI$' represents coherent illumination, $E_O$ is the field distribution of the object, and '*' represents complex conjugation.

Comparing Eqs. (3) and (4), we can find that OMSC-II is phase insensitive and can reflect the object's autocorrelation directly, while OMSC-CI is phase sensitive and the object's autocorrelation should be extracted through square root calculation of an irreversible function. The difference of OMSC between the cases of coherent and incoherent illumination is studied numerically in Figs. 1(d)-1(h), based on the theory of wave optics that is described in the method section. Due to constructive and destructive interference of the coherent wave, the speckle patterns in Fig. 1(f) exhibit higher speckle contrast than those in Fig. 1(d). The OMSC in Fig. 1(g) is noticeably distinct from the autocorrelation of the objects depicted in the lower part of Fig. 1(b). Only by taking the square root operation of Fig. 1(g), the field autocorrelation of the object is revealed in Fig. 1(h), and this is consistent with the expression of Eq. (4). In contrast, for incoherent illumination, the OMSC directly equals to the intensity autocorrelation of the object, i.e., expression of Eq. (3), as demonstrated in Fig. 1(e).

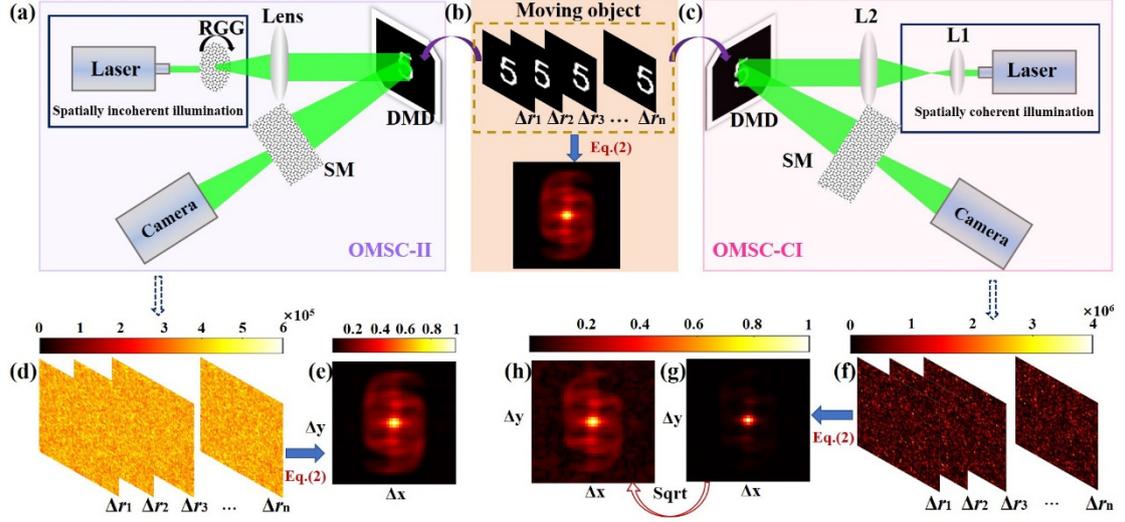

**Fig. 1** Object-motion-based speckle correlation under incoherent illumination and coherent illumination. (a) Schematic of incoherent illumination setup. (b) Images of the moving object and autocorrelation of the object. (c) Schematic of coherent illumination setup. (d), (f) Speckle patterns corresponding to the motion of the object (Fig. 1 (b)) under incoherent illumination and coherent illumination respectively. (e), (g) The cross-correlations (Eq. (2)) between speckle patterns corresponding to (d) and (f) respectively. (h) The square root of Fig. 1 (g). RGG, fast-rotating ground glass. DMD, digital micromirror device. SM, thick scattering medium. L1, Lens 1. L2, Lens 2.

Besides the theoretical analysis, we also conduct experimental verification for the cases of coherent and incoherent illumination, as shown in Supplementary Note 2. Since the hidden object is binary, the object's autocorrelation can be reflected by calculating the OMSC-II and the square root of OMSC-CI, which is the key to reconstructing the hidden object. The experimental results are similar to the conclusion of Fig. 1 and consistent with Eqs. (3) and (4), except that the case of OMSC-CI suffers from a higher noise in the extracted autocorrelation of the object.

Generally, a high signal-to-noise ratio of the autocorrelation pattern is conducive to hidden object reconstruction, especially for objects with high sparsity. Compared to the coherent illumination one, the approach of OMSC-II has lower noise in extracting the autocorrelation from the object-motion-based speckles, which could perform better in reconstructing the hidden object. To further verify this experimentally, objects with different levels of sparsity are considered as shown in the first row of Fig. 2(a), wherein the height of the bars of the hidden objects is gradually increased from 48 pixels to 90 pixels (~10 μm for each pixel) of the DMD to modulate their sparsity. The second row of Fig. 2(a) gives correspondingly ideal autocorrelations of the objects. The third row shows the OMSC under coherent illumination, and the fourth row gives the square root of the third row, which is noticeably similar to the ideal autocorrelation of the object except for background noise. The fifth row of Fig. 2(a) gives the OMSC under incoherent illumination, which is almost the same as the ideal autocorrelations of the objects.

To quantitatively compare the extracted autocorrelation patterns from OMSC-II and OMSC-CI, the root-mean-square error (RMSE) is defined as $RMSE_{II/CI} = 10\log_{10}\left(\left\langle\left|C_{II/CI}^{1/0.5} - A_O\right|^2\right\rangle\right)$, where $C_{II}^1$ denotes the OMSC under incoherent illumination, $C_{CI}^{0.5}$ denotes the square root of OMSC under coherent illumination, and $A_O$ denotes ideal autocorrelation of the object. Figure 2(b) gives the RMSE of OMSC-II (red dotted line) and that of the square root of OMSC-CI (blue dotted line) under ideal numerical simulation conditions (without shot noise, thermal noise, and readout noise). It can be seen that the RMSE under incoherent illumination is lower than that under coherent illumination, which can be explained by the origin of the noise. For OMSC imaging through a thick scattering medium, the intensity distribution of the PSF at different points is considered to be completely uncorrelated. However, when the number of camera pixels is limited, there remains a small nonzero value of correlation between the PSFs since the correlation is calculated by Pearson correlation coefficient, inducing background noise which will be amplified in the case of coherent illumination due to the square root operation. The experimental results given in Fig. 2(c) coincide well with the simulation trend, albeit with some slight fluctuations that may result from experimental noise. In this respect, the approach of OMSC-II performs more effectively in scenes where image reconstructions are conducted from the autocorrelation of the object. The impact of autocorrelation noise on the image reconstruction is illustrated in Supplementary Note 3, which reflects that our proposed approach of OMSC-II can better reconstruct the hidden object through typical phase retrieval algorithms [22-24].

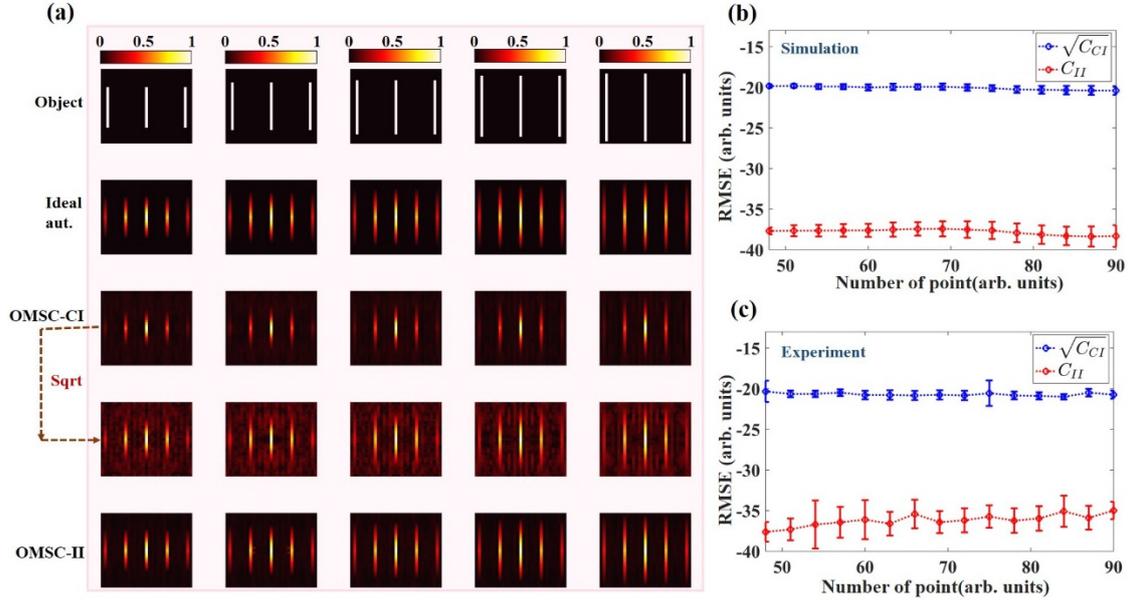

**Fig. 2**. RMSE comparison of objects with different sparsity under incoherent illumination and coherent illumination. (a) The first row contains 5 representative objects, the second row is the corresponding ideal autocorrelations, the third row is the corresponding OMSC-CI, the fourth row is the corresponding square root of OMSC-CI, and the last row is the corresponding OMSC-II. (b) RMSE of OMSC-II (red dotted line) and RMSE of the square root of OMSC-CI (blue dotted line) under ideal simulation conditions. (c) RMSE of OMSC-II (red dotted line) and RMSE of the square root of OMSC-CI (blue dotted line) under experimental conditions. aut., autocorrelation.

## 2. Extrinsic phase modulation: the scene where the object is embedded within a scattering medium.

The above studies focus on the binary object without phase modulation. However, in some practical scenarios, the interested object is generally surrounded by scattering media. As schematically shown in Fig. 3(a), the illuminating beam would first pass through the scattering medium, adding random phase and amplitude modulation to the object when the lighting is spatially coherent. Since the added random phase is independent of the object, we term this extrinsic phase modulation. In this case, the approaches of OMSC-CI and OMSC-II would perform differently.

In the method of OMSC-II, the intensity distribution of the light source on the object plane is still uniform due to the feature of incoherent illumination, making the cross-correlation between speckle patterns remains unaffected (Eq. (3)). In contrast, in the method of OMSC-CI, the interested object will be modulated by a random optical field ($S(r_0)$) when surrounded by scattering media. This makes the cross-correlation between speckle patterns different from the autocorrelation of the object, as given in Eq. (5).

$$C_{CI}(\Delta r) = \left| \frac{\int |S(r_o)|^2 E_O(r_o) E_O^*(r_o + \Delta r) dr_o}{\int |S(r_o)|^2 |E_O(r_o)|^2 dr_o} \right|^2 \tag{5}$$

For incoherent illumination, Fig. 3(b) displays the measured speckles of the object at different displacements, and Fig. 3(c) gives cross-correlation of the speckle patterns, which is almost identical to the ideal autocorrelation (the lower half of Fig. 3(a)). By using the extracted autocorrelation/OMSC, the hidden object can be reconstructed via the well-known phase retrieval algorithm [22-24], as shown in Fig. 3(d). For coherent illumination, the reconstructed image shows little effective information about the interested object (Fig. 3(g)), because the square root of OMSC (the extracted autocorrelation) includes obvious background noise (Fig. 3(f)).

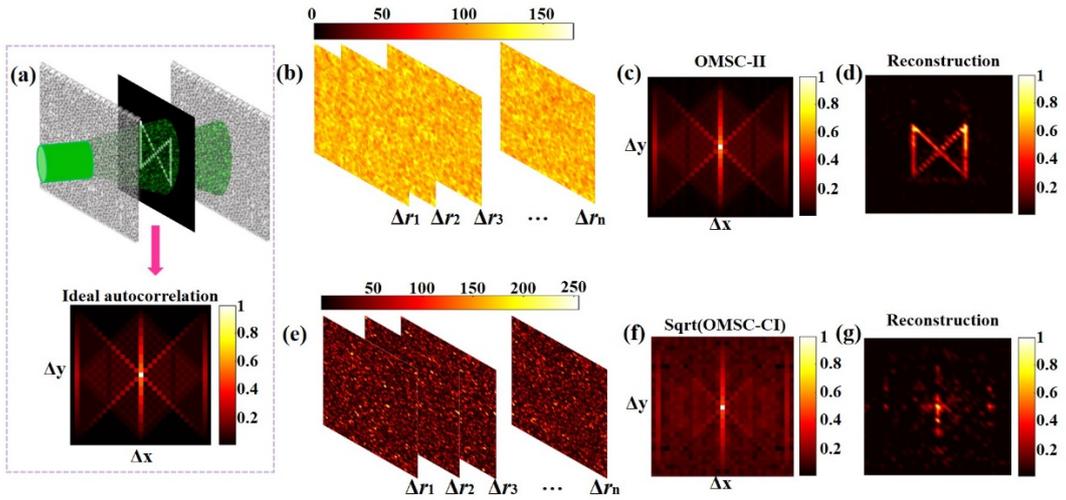

Fig. 3. Scene where objects are embedded within the scattering medium. (a) Schematic of the imaging scene and autocorrelation of the hidden object. (b), (e) Speckles under incoherent illumination and coherent illumination respectively. (c), (f) OMSC-II and the square root of OMSC-CI. (d), (g) Image reconstruction through phase retrieval algorithm for OMSC-II and OMSC-CI respectively.

It is worth noting that there is an ensemble average way to improve the performance of the OMSC-CI approach for the case of objects with extrinsic phase modulation. It resorts to multiple measurements under varied conditions of illumination [20, 21], which has been explained and implemented in Supplementary Note 4. Here, by using the incoherent illumination method, we can directly extract the autocorrelation of hidden objects via the cross-correlation of speckle patterns without ensemble averaging.

## 3. Intrinsic phase modulation: the scene where the object itself causes non-uniform phase modulation.

For some imaging scenarios, the object itself is inhomogeneous, which would impose phase modulation (termed intrinsic phase modulation) even illuminated by an ideal coherent light. In this case, the ensemble average method under coherent illumination is no longer valid because the phase will shift correspondingly with the object and OMSC for coherent illumination (Eq. (4)) is sensitive to phase modulation. In other words, Eq. (4) is a process of coherent superposition and then modulus operation, in which the inhomogeneous phase distribution of the object will lead to the autocorrelation information submerged in noise, resulting in the failure to recover the hidden object through the phase retrieval algorithm. Our approach of OMSC-II can still work, because the OMSC under incoherent illumination (Eq. (3)) is equal to the autocorrelation of the object's intensity which intrinsically ignores any potential random phase modulation on the object.

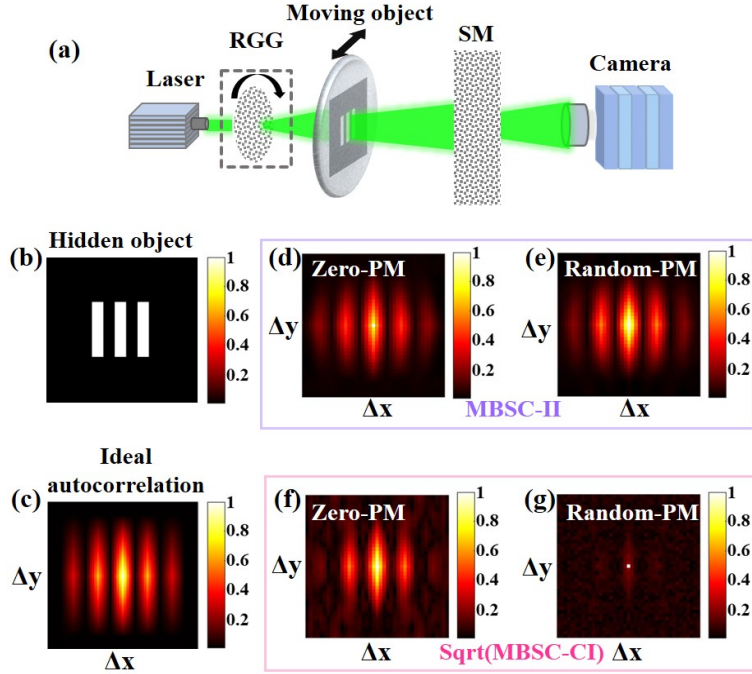

**Fig. 4**. Influence of the intrinsic phase modulation on OMSC-II and OMSC-CI. (a) The experimental setup of OMSC-II (the experimental setup of OMSC-CI is the same except that the RGG is not used). (b) Intensity distribution of the hidden object. (c) Ideal autocorrelation of (b). (d), (f) OMSC-II and OMSC-CI without random phase modulation. (e), (g) OMSC-II and OMSC-CI with random phase modulation. SM, thick scattering medium. Zero-PM, zero phase modulation. RPM, random phase modulation.

Figure 4 gives the experimental demonstration, wherein objects with and without a fixed random phase modulation for the OMSC-II and OMSC-CI approaches are compared. The hidden object consists of a resolution plate and a ground glass that is mounted on a two-dimensional motorized translation stage, as shown in Fig. 4(a). The ground glass is closely attached to the resolution plate to impose a fixed random phase modulation on the object. The hidden object and its real autocorrelation are given in Figs. 4(b) and 4(c) respectively. As a reference, Figs 4(d) and 4(f) give the result when random phase modulation is zero (i.e., the ground glass tightly attached to the resolution plate was removed). Without the pre-added random phase modulation, both two approaches could successfully extract the autocorrelation information of the hidden objects. When the object was introduced with a fixed random phase modulation, the autocorrelation information of the object could still be extracted from OMSC-II (Fig. 4(e)), while the square root of OMSC-CI (Fig. 4(g)) is obviously different from the real autocorrelation (Fig. 4(c)) and can not be used to reconstruct the hidden object. Therefore, although the intrinsic phase information of objects is lost after reconstruction, our approach (OMSC-II) can still acquire the intensity information of objects, which is enough to reflect the object for most scattering imaging applications.

**Discussion**

OMSC is a promising technique that can image auto-moving objects regardless of the shape, material, or thickness of the scattering medium. The core of OMSC is that the autocorrelation of the object can be extracted by calculating the cross-correlation between speckle patterns. For scenes where objects are obscured by thick or complex scattering medium, the PSF has no translation invariance, so the OMSC approach uses a series of speckle patterns related to different displacements of the object to calculate the autocorrelation of the hidden object. For scenes where objects are obscured by a thin scattering medium, the PSF is translation invariant (the speckle patterns translate correspondingly with the object) owing to the existence of memory effects [25]. Therefore, a single speckle pattern is enough for the calculation of OMSC. Actually, the latter scene is called optical memory effect-based speckle-correlated imaging, which has been widely studied and discussed [26-28]. OMSC-II and memory effect-based speckle-correlated imaging are compared in Supplementary Note 5. The latter is suitable for imaging through a thin scattering medium (i.e., the size of an object is within the range of the system's optical memory effect), and OMSC-II is suitable for imaging through a complex scattering medium with negligible optical memory effect.

In conclusion, the influence of spatial coherence is fully investigated for object-motion-based speckle correlation imaging. The incoherent illumination method for object-motion-based imaging through thick scattering media, called OMSC-II, is proposed. Theoretical and experimental results show that the OMSC-II approach can recover the intensity distribution of an object regardless of whether the object is phase-modulated or not and has lower background noise in autocorrelation image, while the OMSC-CI approach is unsuited for reconstruction of phase-modulated objects. This performance difference originates from the fact that OMSC-II corresponds to the autocorrelation of the object's intensity, while OMSC-CI is the squared modulus of field autocorrelation of the object. The proposed method could be extended to other illumination wavelengths, such as X-ray, microwave, near-infrared, and mid-infrared light sources, and facilitates advanced imaging systems including blood cell imaging, star observation, and other motion-based imaging applications.

**Methods**

**Experimental setup**. The experimental system of Fig. 1(a) and Fig. 1(c) consists of a 532 nm laser source (LSR532NL, Ningbo Yuanming Laser Technology Co., Ltd), a rotating ground glass (DG20-600, Thorlabs), optical lenses, a digital micromirror device (DLPLCR65NEVM) serves as a hidden object, and a camera (Zyla 5.5, Andor). In the experiments shown in Figs. 1 to 4, the thick scattering medium between the object and the camera consists of two parallel pieces of ground glass (DG20-220, Thorlabs) that are parallelly placed 3 cm apart, and its optical memory effect range is negligible (details is provided in Supplementary Note 5). The DMD and the camera are synchronized through software. The object of Fig. 4(a) consists of a USAF resolution chart and a piece of ground glass (DG20-600, Thorlabs), which are mounted on a motorized translation stage (NanoMax 342, Thorlabs). The two-dimensional motorized translation stage and camera are synchronized through software. The exposure time of the camera is flexibly adjusted according to the power of illumination.

**Numerical simulation.** The numerical simulation of OMSC is based on the wave optics theorem. For imaging under spatially coherent illumination, the speckle pattern, $I_{CI}$, received on the camera is the squared modulus of the complex field superposition of different *PSFs*, which can be expressed as

$$I_{CI}(r_c; r_o) = \left| \int E_O(r_o) E_{psf}(r_c; r_o) dr_0 \right|^2 \quad (6)$$

For OMSC imaging under spatially incoherent illumination, the speckle pattern, $I_{II}$, received on the camera is the intensity superposition, as described in Eq. (1)

$$I_{II}(r_c; r_o) = \int |E_O(r_o)|^2 |E_{psf}(r_c; r_o)|^2 \, dr_0 \quad (7)$$

The *PSF* ($E_{psf}$) is generated considering that a point propagates from the object plane to the scattering medium plane, and then to the camera plane, which is described based on the Fresnel propagation theorem

$$E_{psf}(r_c; r_o) = \iint_\Sigma e^{\frac{ik}{2u}(x_s^2 + y_s^2)} T(x_s, y_s; r_o) e^{\frac{ik}{2v}[(x_c - x_s)^2 + (y_c - y_s)^2]} dx_s dy_s \quad (8)$$

where the subscripts '*s*' and '*c*' denote the plane of scattering medium and image respectively. *u* and *v* denote the object distance (the distance from the object plane to the scattering medium plane) and the image distance (the distance from the scattering medium plane to the image plane) respectively. *T* represents the random phase modulation from the scattering medium, which is simplified as a pure phase mask. Due to sufficient scattering effect from the thick scattering medium, the spherical wave (the first term in Eq. (8)) generated from the object plane to the scattering medium plane can be omitted. Moreover, the Fraunhofer far-field diffraction distance will be greatly shortened since the optical field from the scattering medium is random [29]. Therefore, Eq. (8) can be simplified as

$$E_{psf}(r_c; r_o) = F\{T(x_s, y_s; r_o)\} \quad (9)$$

where *F* represents the Fourier transform operation. For imaging through a thick scattering medium, the range of memory effect is close to zero, i.e., the *PSFs* are uncorrelated, so the random phase *T* in the numerical simulation is different for different points of the object.

Using Eqs. (6), (7), and (9), the speckles corresponding to different displacements of the object under incoherent and coherent illumination can be generated numerically. Then, the OMSC-II and OMSC-CI are calculated based on Eq. (2).

**Image reconstruction.** After obtaining the autocorrelation image via OMSC, the hidden object is reconstructed through the phase retrieval algorithm. For incoherent illumination, the Fourier amplitude of the object is

$$|F\{I_o\}| = \sqrt{F\{C_{II}\}} \quad (10)$$

where *F* represents the Fourier transform operation. For coherent illumination, the Fourier amplitude of the object is

$$|F\{E_O\}| = \sqrt{F\{\sqrt{C_{CI}}\}} \quad (11)$$

After extracting the Fourier amplitude of the object, the iterative phase-retrieval algorithm (constrained by the combination of hybrid input-output and the error reduction method) can be used to obtain the Fourier phase and finally enables the reconstruction of the hidden object. Detailed information of the phase-retrieval-based reconstruction can be found in Refs. [22-24].

**Funding**

This work was supported by the National Natural Science Foundation of China (NSFC) (11974071, 62375040), the Sichuan Science and Technology Program (2022ZYD0108, 2023JDRC0030), and the Guangdong Basic and Applied Basic Research Foundation under Grant 2020A1515111143.

**Acknowledgments.**

Z. W., M. S., and R. M. developed the theory, and wrote the manuscript. Z. W., A. S., and M. S. performed the experiments. Z. W and W. Z. analyzed the data. R. M., and W. Z. conceived and supervised the project.

**Disclosures.**

The authors declare no competing interests.

**References:**

1. D. Huang, E. A. Swanson, C. P. Lin, J. S. Schuman, W. G. Stinson, W. Chang, M. R. Hee, T. Flotte, K. Gregory, C. A. Puliafito & J. G. Fujimoto. Optical coherence tomography. *Science* **254**, 1178–1181 (1991).
2. L. Wang, P. P. Ho, C. Liu, G. Zhang & R. R. Alfano. Ballistic 2-D imaging through scattering walls using an ultrafast optical Kerr gate. *Science* **253**, 769–771 (1991).
3. W. R. Zipfel, R. M. Williams & W. W. Webb. Nonlinear magic: multiphoton microscopy in the biosciences. *Nat. Biotechnol.* **21**, 1369–1377 (2003).
4. F. Helmchen & W. Denk. Deep tissue two-photon microscopy, *Nat. Methods* **2**, 932–940 (2005).
5. D. Kobat, N. G. Horton & C. Xu, In vivo two-photon microscopy to 1.6-mm depth in mouse cortex. *J. Biomed. Opt.* **16**, 106014 (2011).
6. A. Badon, D. Li, G. Lerosey, A. C. Boccara, M. Fink & A. Aubry. Smart optical coherence tomography for ultra-deep imaging through highly scattering media. *Sci. Adv.* **2**, e1600370 (2016).
7. R. Horstmeyer, H. Ruan & C. Yang. Guidestar-assisted wavefront-shaping methods for focusing light into biological tissue. *Nat. Photon.* **9**, 563–571 (2015).
8. C. M. Woo, Q. Zhao, T. Zhong, H. Li, Z. Yu & P. Lai. Optima efficiency of diffused light focusing through scattering media with iterative wavefront shaping. *APL Photonics* **7**(4), 046109 (2022).
9. I. M. Vellekoop & A. P. Mosk. Focusing coherent light through opaque strongly scattering media. *Opt. Lett.* **32**(16), 2309–11(2007).
10. S. M. Popoff, G. Lerosey, R. Carminati, M. Fink, A. C. Boccara & S. Gigan. Measuring the transmission matrix in optics: an approach to the study and control of light propagation in disordered media. *Phys. Rev. Lett.* **104**,100601 (2010).
11. D. B. Conkey, A. N. Brown, A. M. Caravaca-Aguirre & R. Piestun. Genetic algorithm optimization for focusing through turbid media in noisy environments. *Opt. Express* **20**, 4840–4849 (2012).
12. H. W. Ruan, I. Xu & C.H. Yang. Optical information transmission through complex scattering media with optical-channel-based intensity streaming. *Nat. Commun.* **12**(1), 2411(2021).
13. L. Zhu, F. Soldevila, C. Moretti, A. Arco, A. Boniface, X. Shao, H. B. de Aguiar & S. Gigan, Large field-of-view non-invasive imaging through scattering layers using fluctuating random illumination. *Nat. Commun.* **13**, 1447 (2022).
14. R. Ma, Z. Wang, E. Manuylovich, W. L. Zhang, Y. Zhang, H. Y. Zhu, J. Liu, D. Y. Fan, Y. J. Rao & A. S. L. Gomes. Highly coherent illumination for imaging through opacity. *Opt. Lasers. Eng.* **149**, 106796 (2022).

15. D. J. Wang, S. K. Sahoo, X. Zhu, G. Adamo & C. Dang. Non-invasive super-resolution imaging through dynamic scattering media. *Nat. Commun.* **12**, 3150 (2021).
16. D. Lu, M. Liao, W. He, G. Pedrini, W. Osten & X. Peng. Tracking moving object beyond the optical memory effect. *Opt. Lasers. Eng.* **124**, 105815 (2020).
17. J. A. Newman & K. J. Webb. Imaging optical fields through heavily scattering media. *Phys. Rev. Lett.* **113**, 263903 (2014).
18. K. J. Webb & Q. Luo. Theory of speckle intensity correlations over object position in a heavily scattering random medium. *Phys. Rev. A* **101**, 063827 (2020).
19. J. A. Newman, Q. Luo & K. J. Webb. Imaging hidden objects with spatial speckle intensity correlations over object position. *Phys. Rev. Lett.* **116**, 073902 (2016).
20. Q. Luo, J. A. Newman & K. J. Webb. Motion-based coherent optical imaging in heavily scattering random media. *Opt. Lett.* **44**(11), 2716-2719 (2019).
21. T. Shi, L. Li, H. Cai, X. Zhu, Q. Shi & N. Zheng. Computational imaging of moving objects obscured by a random corridor via speckle correlations. *Nat. Commun.* **13**, 4081 (2022).
22. O. Katz, P. Heidmann, M. Fink & S. Gigan. Non-invasive single-shot imaging through scattering layers and around corners via speckle correlations. *Nat. Photonics* **8**(10), 784-790 (2014).
23. J. R. Fienup. Reconstruction of a complex-valued object from the modulus of its Fourier transform using a support constraint. *J. Opt. Soc. Am. A* **4**, 118–123 (1987).
24. J. Bertolotti, E. G. Van Putten, C. Blum, A. Lagendijk, W. L. Vos & A. P. Mosk. Non-invasive imaging through opaque scattering layers. *Nature* **491**, 232–234 (2012).
25. I. Freund, M. Rosenbluh & S. Feng Memory effects in propagation of optical waves through disordered media. *Phys. Rev. Lett.* **61**, 2328–2331 (1988)
26. Z. Wang, S. Wang, R. Ma, Y. Liu, H. Zhu, Y. Zhang, J. Liu, Y. Mu, Y. Rao & W. Zhang. Near-infrared speckle-illumination imaging based on a multidimensionally disordered fiber laser. *Phys. Rev. Appl.* **18**(2), 024031 (2022).
27. Y. Jauregui-Sánchez, H. Penketh & J. Bertolotti. Tracking moving objects through scattering media via speckle correlations. *Nat. Commun.* **13**(1), 5779 (2022).
28. X. Xie, H. Zhuang, H. He, X. Xu, H. Liang, Y. Liu & J. Zhou. Extended depth-resolved imaging through a thin scattering medium with PSF manipulation. *Sci. Rep.* **8**(1), 4585 (2018).
29. E. Edrei & G. ScarcelliS. Optical imaging through dynamic turbid media using the Fourier-domain shower-curtain effect. *Optica* **3**(1), 71-74 (2016).

# Supplementary material: Incoherent illumination for motion-based imaging through thick scattering medium

**Supplementary Note 1: Theory of OMSC-II and OMSC-CI**

Whether coherent illumination or incoherent illumination, the instantaneous optical field of the observation plane can be expressed as

$$E(r_c; r_o; 2\pi vt) = \int L(r_o; 2\pi vt) E_O(r_o) E_{psf}(r_c; r_o) dr_o \qquad (1.1)$$

Parameter $r$ denotes the transverse spatial coordinate, and the subscripts '$o$' and '$c$' represent the object plane and the camera plane respectively. $L$ is the illumination optical field. $E_O$ is the field distribution of the object. $E_{psf}$ is the random optical field induced by a scattering medium, which represents the mapping relationship between the object plane and the image plane. Because the camera can only respond to the intensity information and the response speed is much slower than the light frequency $v$, the intensity distribution received on the camera can be expressed as

$$I(r_c; r_o) = \left\langle |E(r_c; r_o; 2\pi vt)|^2 \right\rangle_t = \\ \iint \left\langle L(r_o; 2\pi vt) L^*(r_{o'}; 2\pi vt) \right\rangle_t E_O(r_o) E_O^*(r_{o'}) E_{psf}(r_c; r_o) E_{psf}^*(r_c; r_{o'}) dr_o dr_{o'} \qquad (1.2)$$

where $\langle \cdots \rangle_t$ denotes temporal averaging operation. Under uniform coherent illumination

$$\left\langle L(r_o; 2\pi vt) L^*(r_{o'}; 2\pi vt) \right\rangle_t = Cons \qquad (1.3)$$

$$I(r_c; r_o) = \left| \int E_O(r_o) E_{psf}(r_c; r_o) dr_o \right|^2 \qquad (1.4)$$

Under uniform incoherent illumination

$$\left\langle L(r_o; 2\pi vt) L^*(r_{o'}; 2\pi vt) \right\rangle_t = \delta(r_o - r_{o'}) \qquad (1.5)$$

$$I(r_c; r_o) = \int I_O(r_o) I_{psf}(r_c; r_o) dr_o \qquad (1.6)$$

**For OMSC-CI:**

We first derive the relationship between the cross-correlation of speckle patterns and object's auto-correlation under the condition of spatially coherent illumination. The cross-correlation of speckle patterns is calculated as

$$C_{CI}(\Delta r) = \frac{\left\langle \left( I(r_c; r_o) - \langle I(r_c; r_o) \rangle_r \right) \cdot \left( I(r_c; r_o + \Delta r) - \langle I(r_c; r_o + \Delta r) \rangle_r \right) \right\rangle_r}{\sigma_{I(r_c; r_o)} \cdot \sigma_{I(r_c; r_o + \Delta r)}} \qquad (1.7)$$

The subscripts '$CI$' represents the coherent illumination, $\Delta r$ represents the displacement of the object, $\sigma$ is the standard deviation of the speckle, and $\langle \cdots \rangle_r$ denotes spatial averaging operation.

Eq. (1.7) can be simplified as

$$C_{CI}(\Delta r) = \frac{\langle I(r_c; r_o) \cdot I(r_c; r_o + \Delta r) \rangle_r - \langle I(r_c; r_o) \rangle_r \langle I(r_c; r_o + \Delta r) \rangle_r}{\sigma_{I(r_c; r_o)} \cdot \sigma_{I(r_c; r_o + \Delta r)}} \qquad (1.8)$$

When photons are sufficiently scattered (i. e. the phase of scattered light is evenly distributed between $-\pi$ and $\pi$), according to the moment theorem for complex Gaussian [1], the first term of the numerator in Eq. (1.8) can be expanded as

$$\begin{aligned}\langle I(r_c;r_o)\cdot I(r_c;r_o+\Delta r)\rangle_r &= \langle E(r_c;r_o)E^*(r_c;r_o)\cdot E(r_c;r_o+\Delta r)E^*(r_c;r_o+\Delta r)\rangle_r \\ &\approx \langle E(r_c;r_o)E^*(r_c;r_o)\rangle_r \langle E(r_c;r_o+\Delta r)E^*(r_c;r_o+\Delta r)\rangle_r \\ &+\langle E(r_c;r_o)E^*(r_c;r_o+\Delta r)\rangle_r \langle E^*(r_c;r_o)E(r_c;r_o+\Delta r)\rangle_r\end{aligned} \quad (1.9)$$

Substituting (1.9) into (1.8), we have

$$C_{CI}(\Delta r) = \frac{\langle E(r_c;r_o)E^*(r_c;r_o+\Delta r)\rangle_r \langle E^*(r_c;r_o)E(r_c;r_o+\Delta r)\rangle_r}{\sigma_{I(r_c;r_o)}\cdot \sigma_{I(r_c;r_o+\Delta r)}} \quad (1.10)$$

Further, according to Eq. (1.4), the optical field $E(r_c; r_o)$ and $E(r_c; r_{o+}\Delta r)$ can be expressed as

$$E(r_c;r_o) = \int E_{psf}(r_c;r_o)E_O(r_o)dr_o \quad (1.11)$$

$$E(r_c;r_o+\Delta r) = \int E_{psf}(r_c;r_o+\Delta r)E_O(r_o+\Delta r)dr_o \quad (1.12)$$

Substituting (1.11) and (1.12) into (1.10), we have

$$\begin{aligned}&\langle E(r_c;r_o)E^*(r_c;r_o+\Delta r)\rangle_r \\ &= \langle \int E_{psf}(r_c;r_o)E_O(r_o)dr_o \cdot \int E_{psf}^*(r_c;r_o+\Delta r)E_O^*(r_o+\Delta r)dr_o\rangle_r \\ &= \langle |E_{psf}(r_c;r_o)|^2\rangle_r \int E_O(r_o)E_O^*(r_o+\Delta r)dr_o\end{aligned} \quad (1.13)$$

Substituting (1.13) into (1.10), we have

$$\begin{aligned}C_{CI}(\Delta r) &= \frac{\left|\langle |E_{psf}(r_c;r_o)|^2\rangle_r \int E_O(r_o)E_O^*(r_o+\Delta r)dr_o\right|^2}{\sigma_{I(r_c;r_o)}\cdot \sigma_{I(r_c;r_o+\Delta r)}} \\ &= \frac{\langle I_{psf}(r_c;r_o)\rangle_r^2 \left|\int E_O(r_o)E_O^*(r_o+\Delta r)dr_o\right|^2}{\sigma_{I(r_c;r_o)}\cdot \sigma_{I(r_c;r_o+\Delta r)}}\end{aligned} \quad (1.14)$$

Now, consider the denominator

$$\sigma_{I(r_c;r_o)} = \sqrt{\langle (I(r_c;r_o)-\langle I(r_c;r_o)\rangle_r)^2\rangle_r} \quad (1.15)$$

Repeat the process from Eq. (1.7) to Eq. (1.14), we have

$$\sigma_{I(r_c;r_o)} = \langle I_{psf}(r_c;r_o)\rangle_r \int |E_O(r_o)|^2 dr_o \quad (1.16)$$

Similarly,

$$\sigma_{I(r_c;r_o+\Delta r)} = \langle I_{psf}(r_c;r_o+\Delta r)\rangle_r \int |E_O(r_o+\Delta r)|^2 dr_o \quad (1.17)$$

When photons are sufficiently scattered, $\langle I_{psf}\rangle_r$ is approximately the same for different points of the object. Substituting (1.16) and (1.17) into (1.14), we have

$$C_{CI}(\Delta r) = \left|\frac{\int E_O(r_o)E_O^*(r_o+\Delta r)dr_o}{\int |E_O(r_o)|^2 dr_o}\right|^2 \quad (1.18)$$

**For OMSC-II:**

The cross-correlation between speckle patterns under incoherent illumination is calculated as

$$C_{II}(\Delta r) = \frac{\left\langle \left(I(r_c;r_o) - \langle I(r_c;r_o)\rangle_r\right) \cdot \left(I(r_c;r_o+\Delta r) - \langle I(r_c;r_o+\Delta r)\rangle_r\right)\right\rangle_r}{\sigma_{I(r_c;r_o)} \cdot \sigma_{I(r_c;r_o+\Delta r)}} \quad (1.19)$$

The subscript '*II*' represents incoherent illumination. According to (1.6), the intensity distribution of the camera plane can be expressed as

$$I(r_c;r_o) = \int I_{psf}(r_c;r_o) I_O(r_o) dr_o \quad (1.20)$$

$$I(r_c;r_o + \Delta r) = \int I_{psf}(r_c;r_o + \Delta r) I_O(r_o + \Delta r) dr_o \quad (1.21)$$

Substituting (1.20) and (1.21) into (1.19), we have

$$C_{II}(\Delta r) =$$

$$\frac{\left\langle \int \left(I_{psf}(r_c;r_o) - \langle I_{psf}(r_c;r_o)\rangle_r\right) I_O(r_o) dr_o \cdot \int \left(I_{psf}(r_c;r_o+\Delta r) - \langle I_{psf}(r_c;r_o+\Delta r)\rangle_r\right) I_O(r_o+\Delta r) dr_o \right\rangle_r}{\sigma_{I(r_c;r_o)} \cdot \sigma_{I(r_c;r_o+\Delta r)}}$$

$$= \frac{\left\langle \left(I_{psf}(r_c;r_o) - \langle I_{psf}(r_c;r_o)\rangle_r\right)^2\right\rangle_r \int I_O(r_o) I_O(r_o + \Delta r) dr_o}{\sigma_{I(r_c;r_o)} \cdot \sigma_{I(r_c;r_o+\Delta r)}}$$

$$(1.22)$$

When photons are sufficiently scattered, we have [2]

$$\left\langle \left(I_{psf}(r_c;r_o) - \langle I_{psf}(r_c;r_o)\rangle_r\right)^2\right\rangle_r = \langle I_{psf}(r_c;r_o)\rangle_r^2 \quad (1.23)$$

Substituting (1.23) into (1.22)

$$C_{II}(\Delta r) = \frac{\langle I_{psf}(r_c;r_o)\rangle_r^2 \int I_O(r_o) I_O(r_o + \Delta r) dr_o}{\sigma_{I(r_c;r_o)} \cdot \sigma_{I(r_c;r_o+\Delta r)}} \quad (1.24)$$

Repeat the process from Eq. (1.19) to Eq. (1.24), we have

$$\sigma_{I(r_c;r_o)} = \langle I_{psf}(r_c;r_o)\rangle_r \sqrt{\int I_O(r_o)^2 dr_o} \quad (1.25)$$

$$\sigma_{I(r_c;r_o+\Delta r)} = \langle I_{psf}(r_c;r_o+\Delta r)\rangle_r \sqrt{\int I_O(r_o+\Delta r)^2 dr_o} \quad (1.26)$$

Substituting (1.25) and (1.26) into (1.24), we have

$$C_{II}(\Delta r) = \frac{\int I_O(r_o) I_O(r_o + \Delta r) dr_o}{\int I_O(r_o)^2 dr_o} \quad (1.27)$$

In summary (from Eqs. (1.27) and (1.18)), OMSC-II is equal to the autocorrelation of the object's intensity, while the OMSC-CI is equal to the squared modulus of the field autocorrelation of the object.

**Supplementary Note 2: Experimental verification of OMSC-II and OMSC-CI.**

The experimental setup is shown in Fig. 1(a) and 1(c) of the main manuscript, where the thick scattering medium consists of two parallel pieces of ground glass that are parallelly placed 3 cm apart. The pattern of the object and its autocorrelation are shown in Fig. S1(a) and S1(b), respectively. Figures S1(c) and S1(f) show respectively the corresponding speckle patterns of the object under incoherent and coherent illuminations. Then, the cross-correlation between speckle patterns related to different displacements of the object (i.e., object-motion-based speckle correlation, OMSC) is calculated based on Eq. (1.7). Figure S1(d) corresponds to OMSC-II, which is highly similar to the ideal autocorrelation of the object. The regions marked by the white dashed lines in Figs. S1(d) and S1(b) are also compared in Fig. S1(e), which show highly coincident with each other, indicating that the OMSC under incoherent illumination is identical with the autocorrelation of the object and verifying the validity of Eq. (1.27). However, for OMSC-CI, Fig. S1(g) is significantly different from the ideal autocorrelation. After taking a square root operation of Fig. S1(g), the obtained pattern in Fig. S1(h) closely resembles to the ideal autocorrelation of the object. Figure S1(i) compares the regions marked by the white dashed lines in Figs. S1(h) and S1(b). The two curves in Fig. S1(i) are close except for background noise, indicating that the square root of OMSC is consistent with the object's autocorrelation under coherent illumination, as described in Eq. (1.18).

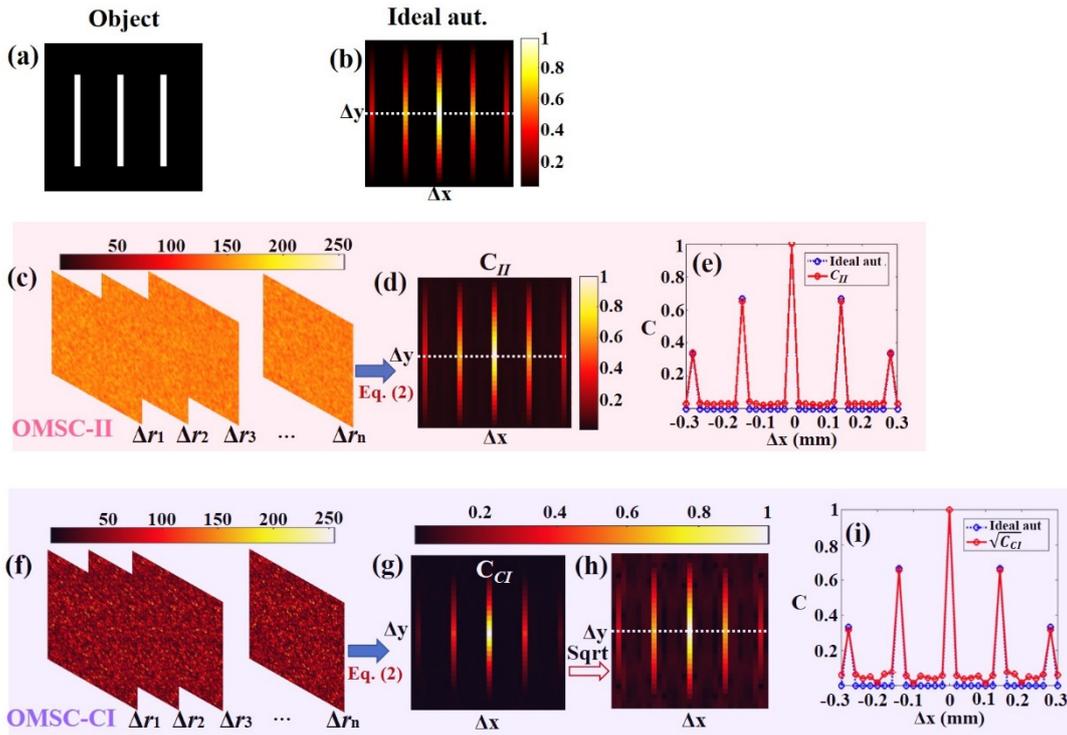

**Fig. S1** Experimental verification of the theory of OMSC-II and comparison between OMSC-II and OMSC-CI. (a) The hidden object. (b) Ideal autocorrelation of (a). (c) Speckles corresponding to different motions of the object under incoherent illumination. (d) Calculated speckle correlation from (c). (e) Comparison between speckle correlation (d) and ideal autocorrelation (b). (f) Speckles correspond to different motions of the object under coherent illumination. (g) Cross-correlation between speckle patterns. (h) The square root of (g). (i) Comparison between (h) and ideal autocorrelation (b). aut., autocorrelation. $C$, correlation.

## Supplementary Note 3: Influence of noise in autocorrelation domain on object reconstruction

According to the Wiener–Khinchin theorem, the object's power spectrum is the Fourier transform amplitude of its autocorrelation, which means the object's Fourier amplitude can be extracted from the autocorrelation (shown in Eq. (3.1)). The missing Fourier phase can be retrieved by the phase-retrieval algorithm and then the hidden object is reconstructed.

$$|F\{O\}| = \sqrt{F\{O \otimes O\}} \quad (3.1)$$

$F$ represents the Fourier transform operation, $\otimes$ represents the autocorrelation operation, and $O$ is the object. If there is noise on the autocorrelation pattern, it will impact the extraction of Fourier amplitudes and disturb the reconstruction of the object. Figure S2 provides a simulation that demonstrates the influence of noise on the phase-retrieval reconstructed image. Figures S2(a) and S2(b) are the reference objects and the corresponding autocorrelation respectively. White Gaussian noise with different intensities is added to the autocorrelation pattern to verify the influence of noise on object reconstruction, as shown in Supplementary Fig. 2(c). From the top downwards, the $RMSE$ of Fig. S2(c) is -20, -25, -30, -35, -40, and -45 dB, wherein $RMSE = 10\log_{10}(\langle |A_N - I_A|^2 \rangle)$, $A_N$ is the abbreviation of autocorrelation with noise and $I_A$ is the abbreviation of ideal autocorrelation. Phase-retrieval reconstructed image corresponding to Fig. S2(c) is shown in Fig. S2(d). It can be seen that when $RMSE$ is greater than -35 dB, the reconstructed image shows little effective information about the object of interest. When $RMSE$ is less than -35 dB, the phase-retrieval reconstructed image is very close to the object.

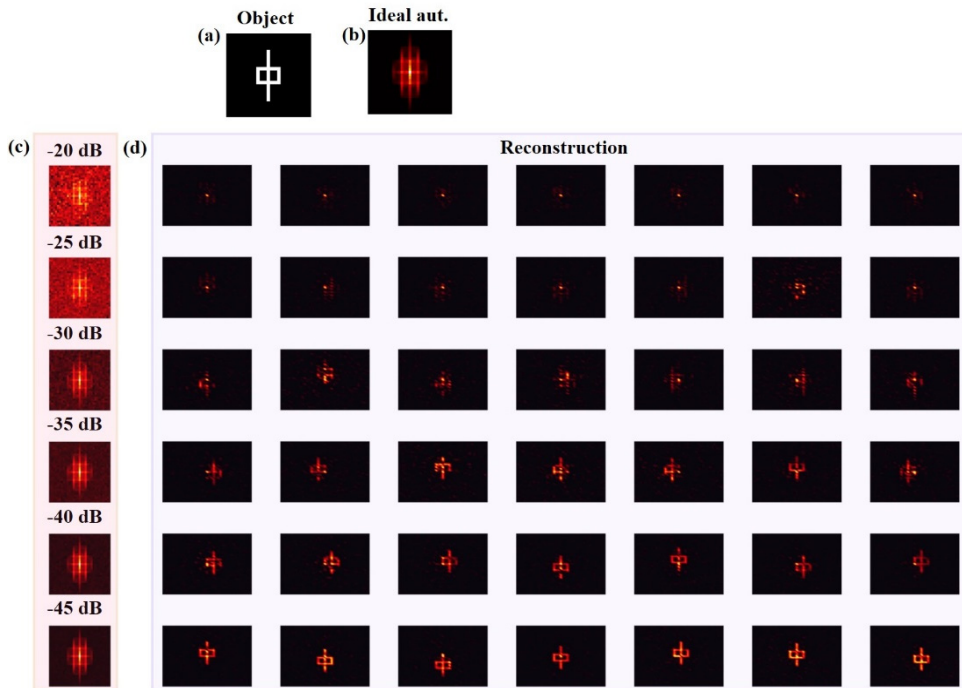

**Fig. S2** Influence of noise on object reconstruction. (a) Object as a reference. (b) Ideal autocorrelation corresponding to (a). (c) Autocorrelation pattern with different noise levels. (d) Phase-retrieval reconstructed image corresponding to (c). aut, autocorrelation.

As verified in the main body and Fig. S1, the extracted autocorrelation image under coherent illumination suffers from higher background noise due to the interference effect and the square root operation. Thus, the proposed OMSC-II method can extract the autocorrelation of hidden objects more precisely, which is useful to reconstruct the object through the phase-retrieval algorithm.

**Supplementary Note 4: Ensemble average suppresses the random optical field**

The ensemble average way can be used to improve the performance of OMSC-CI. Potential ensemble averaging ways can be divided into two categories. One is changing the parameters of coherent illumination, such as lasing wavelength, polarization state, and the incidence angle of light. Another is to select different object positions as a reference to calculate OMSC-CI and then calculate the average to suppress the random noise background.

In this section, we demonstrate that the way of ensemble average improves the cross-correlation of speckle patterns for the coherent illumination approach. Redefine Eq. (1.7) as

$$C_{CI}(\Delta r) = \left\langle \left( I(r_c; r_o) - \left\langle I(r_c; r_o) \right\rangle_r \right) \cdot \left( I(r_c; r_o + \Delta r) - \left\langle I(r_c; r_o + \Delta r) \right\rangle_r \right) \right\rangle_r \quad (4.1)$$

Repeat the process Eq. (1.8) to Eq. (1.13), we have

$$\sqrt{C_{CI}(\Delta r)} = \left\langle I_{psf} \right\rangle_r \int |S(r_o)|^2 E_o(r_o) E_o^*(r_o + \Delta r) dr_o \quad (4.2)$$

When selecting different initial positions as a reference to calculate speckle cross-correlation, we have

$$\sqrt{C_{CI}(\Delta r)} = \left\langle I_{psf} \right\rangle_r \int |S(r_i)|^2 E_o(r_i) E_o^*(r_i + \Delta r) dr_i \quad (4.3)$$

Due to the difference between $S(r_0)$ and $S(r_i)$, superposition between them will suppress the influence of random illumination light fields, as shown in Eq. (4.4).

$$\sqrt{C_{CI}(\Delta r)} = \left\langle I_{psf} \right\rangle_r \int (\frac{1}{n}\sum_{i=1}^{n}|S(r_i)|^2) E_o(r_o) E_o^*(r_o + \Delta r) dr_o$$
$$= C \left\langle I_{psf} \right\rangle_r \int E_o(r_o) E_o^*(r_o + \Delta r) dr_o \quad (4.4)$$

Normalization Eq. (4.4), we have

$$\frac{\sqrt{C_{CI}(\Delta r)}}{\sqrt{C_{CI}(0)}} = \frac{\int E_o(r_o) E_o^*(r_o + \Delta r) dr_o}{\int |E_o(r_o)|^2 dr_o} \quad (4.5)$$

Actually, the process of ensemble averaging is an incoherent superposition to suppress the random light field induced by coherent illumination.

Figures. S3(a) and S3(b) show the ideal autocorrelation of the object and ground truth, respectively. In Figs. S3(c) and S3(d), we give the extracted autocorrelation (the square root of OMSC-CI) and its corresponding phase retrieval reconstructed image without ensemble average, which exhibits a low signal-to-noise ratio. Figures. S3(e), S3(g), and S3(i) are patterns of extracted autocorrelation with 4, 9, and 16 averages, and the reconstructed images are shown in Figs. S3(f), S3(h) and S3(j) respectively. With an increase in the number of average times, the signal-to-noise ratio of the extracted autocorrelation increases, resulting in improved image quality.

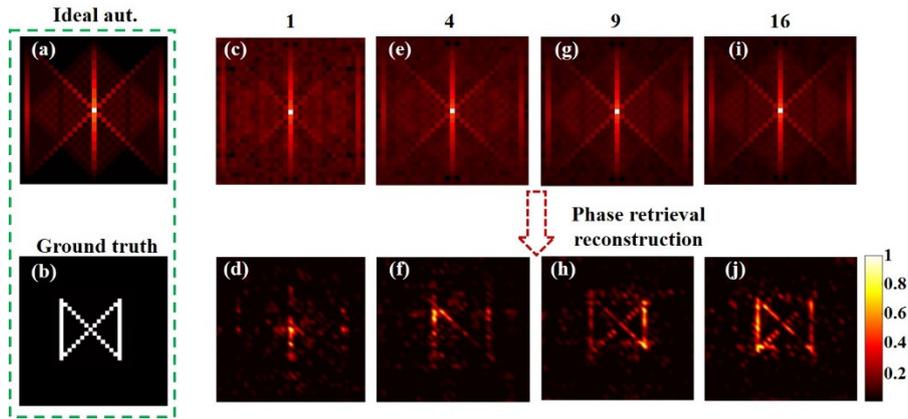

**Fig. S3** Ensemble average suppresses the random optical field. (a) Ideal autocorrelation of the object. (b) Ground truth. (c), (d) Speckle cross-correlation and corresponding reconstructed image without ensemble average. (e), (g), (i) Cross-correlation of speckle patterns with 4, 9, 16 average. (f), (h), (j) Reconstructed image corresponding to (e), (g), (i). aut., autocorrelation.

## Supplementary Note 5: Comparisons between OMSC-II and memory effect-based speckle-correlated imaging

The imaging performance of OMSC-II and optical-memory-effect-based speckle correlated imaging is compared, as shown in Fig. S4. The experimental setup is the same as Fig. 1(a, c) of the main manuscript, where the thick scattering medium consists of two parallel pieces of ground glass that are parallelly placed 3 cm apart. Firstly, one point (size: 5*5 pixels, one pixel is 10.8μm) is loaded onto the DMD and shifted (stepped by 2 pixels) to measure the range of optical memory effect (OME). The range of OME (correlation index decline to 0.5) of a single scattering medium (one piece of ground glass, DG20-220, Thorlabs) is about 23 pixels, as shown in Fig. S4(b). Then, the point is replaced by a letter '*H*' with a size of 15*15 pixels (this size is within the scope of OME) and the corresponding speckle is measured. As is shown in Figs. S4(e, f, g) and Figs. S4(k, l, m), both OMSC-II and optical-memory-effect based speckle correlated imaging can recover object information from speckle patterns. However, the range of OME of a thick scattering medium (two pieces of ground glass, DG20-220, Thorlabs) is less than 2 pixels, as shown in Fig. S4(d). Optical-memory-effect based speckle correlated imaging failed to extract useful information from speckles (shown in Figs. S4(n, o, p)), but OMSC -II still can extract object information (shown in Figs. S4(h, i, j)).

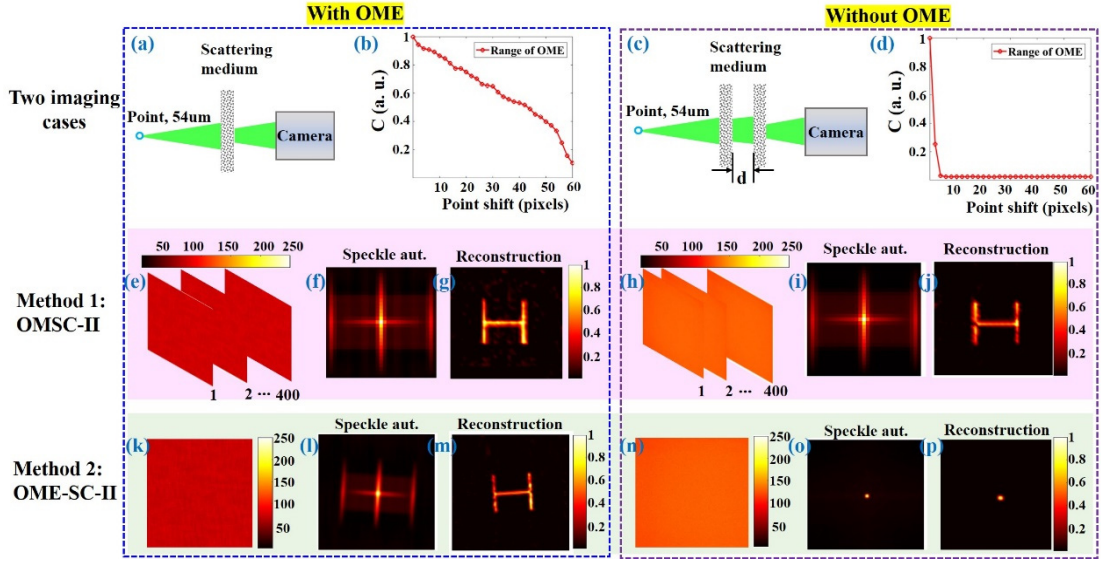

**Fig. S4** Comparisons between OMSC-II and optical-memory-effect based speckle correlated imaging. (a), (c) Imaging model through a thin and thick scattering medium. (b), (d) The range of the optical memory effect of (a) and (c). (e), (f), (g) Speckles, speckle cross-correlation, and the reconstruction of OMSC-II under imaging through a thin scattering medium. (h), (i), (j) Speckles, speckle cross-correlation, and the reconstruction of OMSC-II under imaging through a thick scattering medium. (k), (l), (m) Speckle, speckle auto-correlation, and the reconstruction of optical-memory-effect based imaging under imaging through a thin scattering medium. (n), (o), (p) Speckle, speckle auto-correlation, and the reconstruction of optical-memory-effect based imaging under imaging through a thick scattering medium. C, correlation. d, distance. OME-SC-II, Optical-memory-effect based speckle correlation under incoherent illumination.

**References:**

30. Reed IS. On a moment theorem for complex Gaussian processes. *IRE Transactions on Information Theory* **8**, 194-195 (1962).
31. Goodman JW, Speckle Phenomena in Optics: Theory and Applications. Roberts & Co (2007).